\documentstyle[12pt]{article}
\begin{document}
\baselineskip 20pt
\begin{center}
\baselineskip 20pt
\noindent
{\large \bf THE TOPOLOGICAL QUANTIZATION AND\\
THE BRANCH PROCESS OF THE $(k-1)$%
--DIMENSIONAL TOPOLOGICAL DEFECTS$^*$\footnote{$^*$ Work supported by the National Natural Science
Foundation of P. R. China}}

\vskip 20pt
\normalsize
\noindent
YISHI DUAN, YING JIANG$^\dagger$\footnote{$^\dagger$
Corresponding author; E-mail: itp3@lzu.edu.cn}\\
{\small \it Institute of Theoretical Physics, Lanzhou University,
Lanzhou 730000, P. R. China}\\
\vskip16pt
GUOHONG YANG\\
{\small \it Department of Physics, Fudan University,
Shanghai 200433, P. R. China}

\vskip 36pt
\begin{minipage}{137mm}
\baselineskip 24pt
\normalsize
\noindent
In the light of $\phi $--mapping method and topological current theory, the
topological structure and the topological quantization of arbitrary
dimensional topological defects are obtained under the condition that the
Jacobian $J(\frac \phi v)\neq 0$. When $J(\frac \phi v)=0$, it is shown that
there exist the crucial case of branch process. Based on the implicit
function theorem and the Taylor expansion, we detail the bifurcation of
generalized topological current and find different directions of the
bifurcation. The arbitrary dimensional topological defects are found
splitting or merging at the degenerate point of field function $\vec \phi $ but the
total charge of the topological defects is still unchanged.
\end{minipage}
\end{center}

\vskip 1cm
\noindent
{\bf 1. Introduction}

\vskip 0.5cm

The world of topological defects is amazingly rich and have been the focus
of much attention in many areas of contemporary physics\cite{cos,cosd,gue}.
The importance of the role of defects in understanding a variety of problems
in physics is clear\cite{jan,wang,blamo,cen}. Recently, some physicists
noticed\cite{zyg,neil} that the topological defects are closely related to
the spontaneously broken of $O(m)$ symmetry group to $O(m-1)$ by $m$%
--component order parameter field $\vec \phi $ and pointed out that for $m=1$%
, one has domain walls, $m=2$, strings and $m=3$, monopoles, for $m=4$,
there are textures. But for the lack of a powerful method, the topological
properties are not very clear, the unified theory of describing the
topological properties of all these defect objects is not established yet.

In this paper, in the light of $\phi $--mapping topological current theory%
\cite{11}, a useful method which plays a important role in studying the
topological invariants\cite{lee,li} and the topological structures of
physical systems\cite{14,zhh,4}, we will investigate the topological
quantization and the branch process of arbitrary dimensional topological
defects. We will show that the topological defects are generated from where $%
\vec \phi =0$ and are topological quantized under the condition $J(\frac \phi
v)\neq 0$. While at the zero points of field function $\vec \phi $ where the
corresponding Jacobian determinant $J(\frac \phi v)$ vanishes, the defect
topological current bifurcates and the topological defects split or merge at such
point, this means that the topological defects system is unstable at these
points.

This paper is organized as follows. In section 2, we investigate the
topological quantization of these topological defect and point out that the
topological charges of these defects are the Winding numbers which are
determined by the Hopf indices and the Brouwer degrees of the $\phi $%
--mapping. In section 3, we study the branch process of the defect
topological current at the limit points, bifurcation points and higher
degenerated points systematically by virtue of the $\phi $--mapping theory
and the implicit function theorem.

\vskip 1cm
\noindent
{\bf 2. Topological quantization of topological defects}

\vskip 0.5cm

In our previous papers mentioned above, only the topological current of
point-like particles was discussed. In this paper, in order to study the topological
properties of arbitrary dimensional topological defects, we will extend the
concept to present an arbitrary dimensional generalized topological current.
We consider the $\phi$--mapping as a map between two manifolds, while the
dimensions of the two manifolds are arbitrary. It is an important
generalization of our previous work on topological current and is of great
usefulness to theoretical physics and differential geometry.
        
In $n$--dimensional Riemann manifold $G$ with the metric tensor $g_{\mu \nu }
$ and local coordinates $x^\mu $ ($\mu ,\nu =1,...,n$), a $m$--component
vector order parameter field $\vec \phi (x)$ can be looked upon as a mapping
between the Riemann manifold $G$ and a $m$--dimensional Euclidean space $R^m$%
$$
\phi :\;G\rightarrow R^m,\;\;\;\;\;\;\phi ^a=\phi ^a(x),\;\;\;a=1,...,m. 
$$
The direction field of $\vec \phi (x)$ is generally determined by 
\begin{equation}
\label{0}n^a(x)=\frac{\phi ^a(x)}{||\phi (x)||},\;\;\;\;\;||\phi (x)||=\sqrt{%
\phi ^a(x)\phi ^a(x)}
\end{equation}
with 
\begin{equation}
\label{1}n^a(x)n^a(x)=1.
\end{equation}
It is obviously that $n^a(x)$ is a section of the sphere bundle $S(G)$\cite
{11}. If $n^a(x)$ is a smooth unit vector field without singularities or it
has singularities somewhere but at the point $\vec \phi (x)\neq 0$, from (%
\ref{1}) we have 
\begin{equation}
\label{2}n^a\partial _\mu n^a=0,\;\;\;\;\mu =1,...,n,
\end{equation}
which can be looked upon as a system of $n$ homogeneous linear equations of $%
n^a$ $(a=1,...,m)$ with coefficient matrix $[\partial _\mu n^a]$. The
necessary and sufficient condition that (\ref{2}) has non--trivial solution
for $n^a(x)$ is rank $[\partial _\mu n^a]<m$, i.e. the Jacobian determinants 
\begin{equation}
\label{3}D^{\mu _1\cdot \cdot \cdot \mu _k}(\partial n)=\frac 1{m!}\epsilon
^{\mu _1\cdot \cdot \cdot \mu _k\mu _{k+1}\cdot \cdot \cdot \mu _n}\epsilon
_{a_1\cdot \cdot \cdot a_m}\partial _{\mu _{k+1}}n^{a_1}\cdot \cdot \cdot
\partial _{\mu _n}n^{a_m}
\end{equation}
are equal to zero, where $k=n-m$. While, at the point $\vec \phi =0$, the
above consequences are not held. In short, we have the following relations 
\begin{equation}
\label{4}D^{\mu _1\cdot \cdot \cdot \mu _k}(\partial n)\left\{ 
\begin{array}{cc}
=0, & \;for\;
\vec \phi \neq 0, \\ \neq 0, & \;for\;\vec \phi =0,
\end{array}
\right. 
\end{equation}
which implies $D^{\mu _1\cdot \cdot \cdot \mu _k}(\partial n)$ behaves
itself like a function $\delta (\vec \phi )$. So we are focussed on the
zeroes of $\phi ^a(x)$.

Suppose that the vector field $\vec \phi (x)$
possesses $l$ isolated zeroes, according to the implicit function theorem\cite{golsat}%
, when the zeroes are regular points of $\phi $--mapping at which the rank
of the Jacobian matrix $[\partial _\mu \phi ^a]$ is $m$, the solutions of $%
\vec \phi =0$ can be expressed parameterizedly by 
\begin{equation}
\label{5}x^\mu =z_i^\mu (u^1,\cdot \cdot \cdot ,u^k),\;\;\;\;i=1,...,l,
\end{equation}
where the subscript $i$ represents the $i$--th solution and the parameters $%
u^I$ ($I=1,...,k$) span a $k$--dimensional submanifold with the metric
tensor $g_{IJ}=g_{\mu \nu }\frac{\partial x^\mu }{\partial u^I}\frac{%
\partial x^\nu }{\partial u^J}$ which is called the $i$--th singular
submanifold $N_i$ in the Riemannian manifold $G$ corresponding to the $\phi $%
--mapping. For each singular manifold $N_i$, we can define a normal
submanifold $M_i$ in $G$ which is spanned by the parameters $v^A$ with the
metric tensor $g_{AB}=g_{\mu \nu }\frac{\partial x^\mu }{\partial v^A}\frac{%
\partial x^\nu }{\partial v^B}$ $(A,B=1,...,m)$, and the intersection point of
$M_i$ and $N_i$ is denoted by $p_i$ which can be expressed parameterizedly
by $v^A=p_i^A$. In fact, in the words of differential topology, $M_i$ is
transversal to $N_i$ at the point $p_i$, i.e.
$$
T_{p_i}(G)=T_{p_i}(M_i)+T_{p_i}(N_i).
$$
By virtue of the implicit function
theorem, it should be held true that, at the regular point $p_i$, the Jacobian
matrices $J(\frac \phi v)$ satisfies 
\begin{equation}
\label{nonzero}J(\frac \phi v)=\frac{D(\phi ^1,\cdot \cdot \cdot ,\phi ^m)}{%
D(v^1,\cdot \cdot \cdot ,v^m)}\neq 0.
\end{equation}

In the following, we will induce a rank--$k$ topological current through the
integration of $D^{\mu _1\cdot \cdot \cdot \mu _k}(\partial n)$ in (\ref{3})
on $M_i$. As is well known, the generalized Winding Number\cite{19} has been
given by the Gauss map $n:\partial \Sigma _i\rightarrow S^{m-1}$%
\begin{equation}
\label{6}W_i=\frac 1{A(S^{m-1})(m-1)!}\int_{\partial \Sigma
_i}n^{*}(\epsilon _{a_1\cdot \cdot \cdot a_m}n^{a_1}dn^{a_2}\wedge \cdot
\cdot \cdot \wedge dn^{a_m}) 
\end{equation}
where%
$$
A(S^{m-1})=\frac{2\pi ^{m/2}}{\Gamma (m/2)} 
$$
is the area of ($m-1$)--dimensional unit sphere $S^{m-1}$, $n^{*}$ denotes
the pull back of map $n$ and $\partial \Sigma _i$ the boundary of a
neighborhood $\Sigma _i$ of $p_i$ on $M_i$ with $p_i\notin \partial \Sigma
_i $, $\Sigma _i\cap \Sigma _j=\emptyset $. The generalized Winding Numbers $%
W_i $ can also be rewritten as%
$$
W_i=\frac 1{A(S^{m-1})(m-1)!}\int_{n[\partial \Sigma _i]}\epsilon _{a_1\cdot
\cdot \cdot a_m}n^{a_1}dn^{a_2}\wedge \cdot \cdot \cdot \wedge dn^{a_m} 
$$
which means that, when the point $x^\mu $ or $v^A$ covers $\partial \Sigma
_i $ once, the unit vector $n^a$ will cover a region $n[\partial \Sigma _i]$
whose area is $W_i$ times of $A(S^{m-1})$, i.e. the unit vector $n^a$ will
cover the unit sphere $S^{m-1}$ $W_i$ times. From the above equation, one
can deduce that%
$$
W_i=\frac 1{A(S^{m-1})(m-1)!}\int_{\partial M_i}\epsilon _{a_1\cdot \cdot
\cdot a_m}n^{a_1}\partial _{\mu _{k+2}}n^{a_2}\cdot \cdot \cdot \partial
_{\mu _n}n^{a_m}dx^{\mu _{k+2}}\wedge \cdot \cdot \cdot \wedge dx^{\mu _n} 
$$
$$
=\frac 1{A(S^{m-1})(m-1)!}\int_{M_i}\epsilon _{a_1\cdot \cdot \cdot
a_m}\partial _{\mu _{k+1}}n^{a_1}\partial _{\mu _{k+2}}n^{a_2}\cdot \cdot
\cdot \partial _{\mu _n}n^{a_m}dx^{\mu _{k+1}}\wedge \cdot \cdot \cdot
\wedge dx^{\mu _n} 
$$
$$
=\frac 1{A(S^{m-1})(m-1)!}\int_{M_i}\frac 1{k!}\frac 1{\sqrt{g_x}}\epsilon
^{\mu _1\cdot \cdot \cdot \mu _k\mu _{k+1}\cdot \cdot \cdot \mu _n}\epsilon
_{a_1\cdot \cdot \cdot a_m}\partial _{\mu _{k+1}}n^{a_1}\partial _{\mu
_{k+2}}n^{a_2}\cdot \cdot \cdot \partial _{\mu _n}n^{a_m}d\sigma _{\mu
_1\cdot \cdot \cdot \mu _k} 
$$
\begin{equation}
\label{7}=\frac 1{A(S^{m-1})(m-1)!}\int_{M_i}\frac 1{k!}\frac{m!}{\sqrt{g_x}}%
D^{\mu _1\cdot \cdot \cdot \mu _k}(\partial n)d\sigma _{\mu _1\cdot \cdot
\cdot \mu _k}, 
\end{equation}
where $d\sigma _{\mu _1\cdot \cdot \cdot \mu _k}$ is the invariant surface
element of $M_i$ and $g_x=\det (g_{\mu \nu })$.

From the above discussions, especially the expressions (\ref{3}), (\ref{4})
and (\ref{7}), we can induce a generalized topological current $j^{\mu
_1\cdot \cdot \cdot \mu _k}$ which does not vanish only at the zeroes of
order parameter field $\vec \phi (x)$, and is exactly corresponding to the
generalized Winding Number, 
\begin{equation}
\label{8}j^{\mu _1\cdot \cdot \cdot \mu _k}=\frac 1{A(S^{m-1})(m-1)!\sqrt{g_x%
}}\epsilon ^{\mu _1\cdot \cdot \cdot \mu _k\mu _{k+1}\cdot \cdot \cdot \mu
_n}\epsilon _{a_1\cdot \cdot \cdot a_m}\partial _{\mu _{k+1}}n^{a_1}\partial
_{\mu _{k+2}}n^{a_2}\cdot \cdot \cdot \partial _{\mu _n}n^{a_m}.
\end{equation}
Obviously this tensor current is identically conserved, i.e.%
$$
\nabla _{\mu _i}j^{\mu _1\cdot \cdot \cdot \mu _k}=0,\;\;\;i=1,...,k. 
$$
It is easy to see that $j^{\mu _1\cdot \cdot \cdot \mu _k}$ are completely
antisymmetric tensors.

By making use of the $\phi $--mapping theory, we will study the global
property of the generalized topological current $j^{\mu _1\cdot \cdot \cdot
\mu _k}$ on the whole manifold $G$ and conclude that $j^{\mu _1\cdot \cdot
\cdot \mu _k}$ behaves itself like the generalized function $\delta (\vec 
\phi )$. From (\ref{0}) we have%
$$
\partial _\mu n^a=\frac 1{||\phi ||}\partial _\mu \phi ^a+\phi ^a\partial
_\mu (\frac 1{||\phi ||}),\;\;\;\frac \partial {\partial \phi ^a}(\frac 1{%
||\phi ||})=-\frac{\phi ^a}{||\phi ||^3} 
$$
which should be looked upon as generalized functions\cite{gelfand}. Using
these expressions the generalized topological current (\ref{8}) can be
rewritten as%
\begin{eqnarray}
j^{\mu _1\cdot \cdot \cdot \mu _k}&=&C_m\frac 1{\sqrt{g_x}}\epsilon ^{\mu
_1\cdot \cdot \cdot \mu _k\mu _{k+1}\cdot \cdot \cdot \mu _n}\epsilon
_{a_1\cdot \cdot \cdot a_m}\nonumber \\
& &\cdot \partial _{\mu _{k+1}}\phi ^a\partial _{\mu _{k+2}}\phi ^{a_2}\cdots
\partial _{\mu _n}\phi ^{a_m}\frac \partial {\partial \phi ^a}\frac \partial
{\partial \phi ^{a_1}}(G_m(||\phi ||)),\;\;\;\;\;m>2. 
\end{eqnarray}
where $C_m$ is a constant%
$$
C_m=\left\{ 
\begin{array}{cc}
-\frac 1{A(S^{m-1})(m-2)(m-1)!}, & \;\;\;\;m>2 \\ 
\frac 1{2\pi }, & \;\;\;\;m=2
\end{array}
,\right.  
$$
and $G_m(||\phi ||)$ is a generalized function%
$$
G_m(||\phi ||)=\left\{ 
\begin{array}{ccc}
\frac 1{||\phi ||^{m-2}} & ,\;\;\; & m>2 \\ 
\ln ||\phi || & ,\;\;\; & m=2
\end{array}
.\right.  
$$
Defining general Jacobians $J^{\mu _1\cdot \cdot \cdot \mu _k}(\frac \phi x)$
as following%
$$
\epsilon ^{a_1\cdot \cdot \cdot a_m}J^{\mu _1\cdot \cdot \cdot \mu _k}(\frac 
\phi x)=\epsilon ^{\mu _1\cdot \cdot \cdot \mu _k\mu _{k+1}\cdot \cdot \cdot
\mu _n}\partial _{\mu _{k+1}}\phi ^{a_1}\partial _{\mu _{k+2}}\phi
^{a_2}\cdot \cdot \cdot \partial _{\mu _n}\phi ^{a_m} 
$$
and by making use of the $m$--dimensional Laplacian Green function relation%
\cite{11}%
$$
\Delta _\phi (\frac 1{||\phi ||^{m-2}})=-\frac{4\pi ^{m/2}}{\Gamma (\frac m2%
-1)}\delta (\vec \phi ) 
$$
where $\Delta _\phi =(\frac{\partial ^2}{\partial \phi ^a\partial \phi ^a})$
is the $m$--dimensional Laplacian operator in $\phi $--space, we do obtain
the $\delta $--function like topological current rigorously 
\begin{equation}
\label{10}j^{\mu _1\cdot \cdot \cdot \mu _k}=\frac 1{\sqrt{g_x}}\delta (\vec 
\phi )J^{\mu _1\cdot \cdot \cdot \mu _k}(\frac \phi x).
\end{equation}
We find that $j^{\mu _1\cdot \cdot \cdot \mu _k}\neq 0$ only when $\vec \phi
=0$, which is just the singularity of $j^{\mu _1\cdot \cdot \cdot \mu _k}$.
In detail, the Kernel of the $\phi $--mapping is the singularities of the
topological tensor current $j^{\mu _1\cdot \cdot \cdot \mu _k}$ in $G$. We
think that this is the essential of the topological tensor current theory
and $\phi $--mapping is the key to study this theory.

To investigate the topological properties of the generalized topological
current, we should study the total expansion of the $\delta$--function
$\delta (\vec \phi)$. As is well known\cite{17}, the $\delta $--function
$\delta (N_i)$ in curved space-time on a submanifold $N_i$ is
\begin{equation}
\label{m}\delta (N_i)=\int_{N_i}\frac 1{\sqrt{g_x}}\delta ^n(\vec x-\vec z%
_i(u^1,u^2))\sqrt{g_u}d^ku, 
\end{equation}
and, by analogy with the procedure of deducing $\delta (f(x))$, since 
\begin{equation}
\delta (\vec \phi )=\left\{ 
\begin{array}{cc}
+\infty , & for\; 
\vec \phi (x)=0 \\ 0, & for\;\vec \phi (x)\neq 0 
\end{array}
\right. =\left\{ 
\begin{array}{cc}
+\infty , & for\;x\in N_i \\ 
0, & for\;x\notin N_i 
\end{array}
\right. , 
\end{equation}
we can expand the $\delta $--function $\delta (\vec \phi )$ as 
\begin{equation}
\label{delta}\delta (\vec \phi )=\sum_{i=1}^lc_i\delta (N_i), 
\end{equation}
where the coefficients $c_i$ must be positive, i.e. $c_i=\mid c_i\mid $.
From the definition of $W_i$ in (\ref{6}), the Winding number can also be
rewritten in terms of the parameters $v^A$ of $M_i$ as 
$$
W_i=\frac 1{2\pi }\int_{\Sigma _i}\epsilon ^{A_1\cdot \cdot \cdot
A_m}\epsilon _{a_1\cdot \cdot \cdot a_m}\partial _{A_1}n^{a_1}\cdot \cdot
\cdot \partial _{A_m}n^{a_m}d^mv, 
$$
Then, by duplicating the above process, we have 
\begin{equation}
\label{W}W_i=\int_{\Sigma _i}\delta (\vec \phi )J(\frac \phi v)d^mv, 
\end{equation}
Substituting (\ref{delta}) into (\ref{W}), and considering that only one $%
p_i\in \Sigma _i$, we can get 
\begin{equation}
W_i=\int_{\Sigma _i}c_i\delta (N_i)J(\frac \phi v)d^mv=\int_{\Sigma
_i}\int_{N_i}c_i\frac 1{\sqrt{g_x}\sqrt{g_v}}\delta ^n(\vec x-\vec z%
_i(u^1,u^2))J(\frac \phi v)\sqrt{g_u}d^ku\sqrt{g_v}d^mv. 
\end{equation}
where $g_v=\det (g_{AB})$. Because $\sqrt{g_u}\sqrt{g_v}d^kud^mv$ is the
invariant volume element of the Product manifold $M_i\times N_i$, so it can
be rewritten as $\sqrt{g_x}d^nx$. Thus, by calculating the integral and with
positivity of $c_i$, we get 
\begin{equation}
c_i=\frac{\beta _i\sqrt{g_v}}{\mid J(\frac \phi v)_{p_i}\mid }=\frac{\beta
_i\eta _i\sqrt{g_v}}{J(\frac \phi v)_{p_i}}, 
\end{equation}
where $\beta _i=|W_i|$ is a positive integer called the Hopf index\cite{20}
of $\phi $-mapping on $M_i,$ it means that when the point $v$ covers the
neighborhood of the zero point $p_i$ once, the function $\vec \phi $ covers
the corresponding region in $\vec \phi $-space $\beta _i$ times, and $\eta
_i=signJ(\frac \phi v)_{p_i}=\pm 1$ is the Brouwer degree of $\phi $-mapping%
\cite{20}. Substituting this expression of $c_i$ and (\ref{delta}) into (\ref
{10}), we gain the total expansion of the rank--$k$ topological current 
$$
j^{\mu _1\cdot \cdot \cdot \mu _k}=\frac 1{\sqrt{g_x}}\sum_{i=1}^l\frac{%
\beta _i\eta _i\sqrt{g_v}}{J(\frac \phi v)|_{p_i}}\delta (N_i)J^{\mu _1\cdot
\cdot \cdot \mu _k}(\frac \phi x). 
$$
or in terms of parameters $y^{A^{^{\prime }}}=(v^1,\cdot \cdot \cdot
,v^m,u^1,\cdot \cdot \cdot ,u^k)$%
\begin{equation}
j^{A_1^{^{\prime }}\cdot \cdot \cdot A_k^{^{\prime }}}=\frac 1{\sqrt{g_y}}%
\sum_{i=1}^l\frac{\beta _i\eta _i\sqrt{g_v}}{J(\frac \phi v)|_{p_i}}\delta
(N_i)J^{A_1^{^{\prime }}\cdot \cdot \cdot A_k^{^{\prime }}}(\frac \phi y).
\end{equation}
From the above equation, we conclude that the inner structure of $j^{\mu
_1\cdot \cdot \cdot \mu _k}$ or $j^{A_1^{^{\prime }}\cdot \cdot \cdot
A_k^{^{\prime }}}$ is labelled by the total expansion of $\delta (\vec \phi
) $, which includes the topological information $\beta _i$ and $\eta _i.$

It is obvious that, in (\ref{5}), when $u^1$ and$\,u^I(I=2,...,k)$ are taken
to be time-like evolution parameter and space-like parameters, respectively,
the inner structure of $j^{\mu _1\cdot \cdot \cdot \mu _k}$ or $%
j^{A_1^{^{\prime }}\cdot \cdot \cdot A_k^{^{\prime }}}$ just represents $l$
$(k-1)$--dimensional topological defects moving in the $n$--dimensional Riemann
manifold $G$. The $k$-dimensional singular submanifolds $N_i\,\,(i=1,\cdot
\cdot \cdot l)$ are their world sheets. Here we see that the defects are
generated from where $\vec \phi =0$ and, the Hopf indices $\beta _i$ and
Brouwer degree $\eta _i$ classify these defects. In detail, the Hopf indices 
$\beta _i$ characterize the absolute values of the topological quantization
and the Brouwer degrees $\eta _i=+1$ correspond to defects while $\eta _i=-1$
to antidefects. It must be pointed that the relationship between the zero
points of the $m$--dimensional order parameter field $\vec \phi $ and the
space position of these topological defects is distinct and clear and it is
obtained rigorously without tie on any concrete model or hypothesis.
Furthermore, for the first time we gain the topological charges of these
defects which are determined by the Winding numbers of the $\phi $--mapping.

\vskip 1cm
\noindent
{\bf 3. The branch processes of the topological defects}

\vskip 0.5cm

In this section, we will discuss the branch processes of these topological
defects. In order to simplify our study, we select the parameter $u^1$ as
the time--like evolution parameter $t$, and let the space--like parameters $%
u^I=\sigma ^I\;(I=2,...,k)$ be fixed. In this case, the Jacobian matrices $%
J^{A_1^{^{\prime }}\cdot \cdot \cdot A_k^{^{\prime }}}(\frac \phi y)$ are
reduced to%
$$
J^{AI_1\cdot \cdot \cdot I_{k-1}}(\frac \phi y)\equiv J^A(\frac \phi y%
),\;\;\;\;J^{ABI_1\cdot \cdot \cdot I_{k-2}}(\frac \phi y)=0,\;\;\;%
\;J^{(m+1)\cdot \cdot \cdot n}(\frac \phi y)=J(\frac \phi v), 
$$
\begin{equation}
A,B=1,...,(m+1),\;\;\;\;I_j=m+2,...,n, 
\end{equation}
for $y^A=v^A\;(A\leq m),\;y^{m+1}=t,\;y^{m+I}=\sigma ^I\;(I\geq 2)$. In the
above section, we have studied the topological property of the topological
defects in the case that the vector order parameter $\vec \phi $ only
consists of regular points, i.e. (\ref{nonzero}) is hold true. However, when
this condition fails, the above results will change in some way. It often
happens when the zeros of $\vec \phi $ include some branch points, which
lead to the branch process of topological current. The branch points are
determined by the $m+1$ equations 
\begin{equation}
\label{phia}\phi ^a(v^1,\cdots ,v^m,t,\vec \sigma )=0,\;\;\;a=1,...,m 
\end{equation}
and 
\begin{equation}
\label{zero}\phi ^{m+1}(v^1,\cdots ,v^m,t,\vec \sigma )\equiv J(\frac \phi v%
)=0 
\end{equation}
for the fixed $\vec \sigma $. and they are denoted as $(t^{*},p_i)$. In $%
\phi $--mapping theory usually there are two kinds of branch points, namely
the limit points and bifurcation points\cite{27}, satisfying 
\begin{equation}
\label{nonzero1}J^1(\frac \phi y)|_{(t^{*},p_i)}\neq 0 
\end{equation}
and 
\begin{equation}
\label{zero1}J^1(\frac \phi y)|_{(t^{*},p_i)}=0, 
\end{equation}
respectively. In the following, we assume that the branch points $%
(t^{*},p_i) $ of $\phi $--mapping have been found.

\vskip 0.5cm

{\bf A. Branch process at the limit point}

\vskip 0.5cm
In order to use the theorem of implicit function to study the branch process
of topological defects at the limit point, we use the Jacobian $J^1(\frac 
\phi y)$ instead of $J(\frac \phi v)$ to discuss the problem. In fact, this
means that we have replaced the parameter $t$ by $v^1$. For clarity we
rewrite the problem as 
\begin{equation}
\label{101}\phi ^a(t,v^2,\cdots ,v^m,v^1,\vec \sigma )=0,\;\;\;\;%
\;a=1,...,m. 
\end{equation}
Then, taking account of the condition(\ref{nonzero1}) and using the implicit
function theorem, we have an unique solution of the equations (\ref{101}) in
the neighborhood of the limit point $(t^{*},p_i)$%
\begin{equation}
\label{102}t=t(v^1,\vec \sigma ),\;\;\;\;v^i=v^i(v^1,\vec \sigma
),\;\;\;\;i=2,3,...,m 
\end{equation}
with $t^{*}=t(p_i^1,\vec \sigma )$. In order to show the behavior of the
defects at the limit points, we will investigate the Taylor expansion of (%
\ref{102}) in the neighborhood of $(t^{*},p_i)$. In the present case, from (%
\ref{nonzero1}) and (\ref{zero}), we get
\begin{equation}
\label{bifa18}
\frac{dv^1}{dt}|_{(t^{*},p_i)}=\frac{J^1(\frac \phi y)}{J(\frac \phi y)}%
|_{(t^{*},p_i)}=\infty , 
\end{equation}
i.e.%
$$
\frac{dt}{dv^1}|_{(t^{*},p_i)}=0. 
$$
Then we have the Taylor expansion of (\ref{102}) at the point $(t^{*},p_i)$%
$$
t=t(p_i,\vec \sigma )+\frac{dt}{dv^1}|_{(t^{*},p_i)}(v^1-p_i^1)+\frac 12%
\frac{d^2t}{(dv^1)^2}|_{(t^{*},p_i)}(x^1-p_i^1)^2 
$$
$$
=t^{*}+\frac 12\frac{d^2t}{(dv^1)^2}|_{(t^{*},p_i)}(v^1-p_i^1)^2. 
$$
Therefore 
\begin{equation}
\label{103}t-t^{*}=\frac 12\frac{d^2t}{(dv^1)^2}|_{(t^{*},p_i)}(v^1-p_i^1)^2 
\end{equation}
which is a parabola in the $v^1$---$t$ plane. From (\ref{103}), we can
obtain the two solutions $v_{(1)}^1(t,\vec \sigma )$ and $v_{(2)}^1(t,\vec 
\sigma )$, which give the branch solutions of the system (\ref{phia}) at the
limit point. If $\frac{d^2t}{(dv^1)^2}|_{(t^{*},p_i)}>0$, we have the branch
solutions for $t>t^{*}$, otherwise, we have the branch solutions for $%
t<t^{*} $. The former is related to the origin of the topological defects at
the limit points. Since the topological current of the topological defects
is identically conserved, the topological quantum numbers of these two
generated defects must be opposite at the limit point, i.e. $\beta _1\eta
_1+\beta _2\eta _2=0$. In fact, these two cases are just related to the
generation and annihilation of defect-antidefect pair. The result (\ref{bifa18})
agrees with that obtained by Bray \cite{Bray} who had a scaling argument associated
with point defects final annihilation which leads to large velocity tail.

\vskip0.5cm
{\bf B. Branch process at the bifurcation point}

\vskip0.5cm
In the following, let us consider the case (\ref{zero1}), in which the
restrictions of the system (\ref{phia}) at the bifurcation point $%
(t^{*},p_i) $ are 
\begin{equation}
\label{104}J(\frac \phi v)|_{(t^{*},p_i)}=0,\;\;\;J(\frac \phi v%
)|_{(t^{*},p_i)}=0. 
\end{equation}
These two restrictive conditions will lead to an important fact that the
dependency relationship between $t$ and $v^1$ is not unique in the
neighborhood of the bifurcation point $(t^{*},p_i).$ In fact, we have 
\begin{equation}
\label{105}\frac{dv^1}{dt}|_{(t^{*},p_i)}=\frac{J^1(\frac \phi y)}{J(\frac 
\phi v)}|_{(t^{*},p_i)} 
\end{equation}
which under the restraint (\ref{104}) directly shows that the tangential
direction of the integral curve of equation (\ref{105}) is indefinite at the
point $(t^{*},p_i)$. Hence, (\ref{105}) does not satisfy the conditions of
the existence and uniqueness theorem of the solution of a differential
equation. This is why the very point $(t^{*},\vec z_i)$ is called the
bifurcation point of the system (\ref{phia}).

In the following, we will find a simple way to search for the different
directions of all branch curves at the bifurcation point. As assumed that
the bifurcation point $(t^{*},p_i)$ has been found from (\ref{phia}) and (%
\ref{zero}), the following calculations are all conducted at the value $%
(t^{*},p_i)$. As we have mentioned above, at the bifurcation point $%
(t^{*},p_i)$, the rank of the Jacobian matrix $[\frac{\partial \phi }{%
\partial v}]$ is smaller than $m$. In order to derive the calculating
method, we consider the rank of the Jacobian matrix $[\frac{\partial \phi }{%
\partial v}]$ is $m-1$. The case of a more smaller rank will be discussed in
next subsection. Suppose that one of the $(m-1)\times (m-1)$ submatrix $J_1(%
\frac \phi v)$ of the Jacobian matrix $[\frac{\partial \phi }{\partial v}]$
is 
\begin{equation}
\label{106}J_1(\frac \phi v)=\left( 
\begin{array}{cccc}
\phi _2^1 & \phi _3^1 & \cdots & \phi _m^1 \\ 
\phi _2^2 & \phi _3^2 & \cdots & \phi _m^2 \\ 
\vdots & \vdots & \ddots & \vdots \\ 
\phi _2^{m-1} & \phi _3^{m-1} & \cdots & \phi _m^{m-1} 
\end{array}
\right) 
\end{equation}
and its determinant $\det J_1(\frac \phi v)$ does not vanish at the point $%
(t^{*},p_i)$ (otherwise, we have to rearrange the equations of (\ref{phia}%
)), where $\phi _A^a$ stands for $(\partial \phi ^a/\partial v^A)$ $%
(a=1,...,m-1;\;A=2,...,m)$. By means of the implicit function theorem we
obtain one and only one functional relationship in the neighborhood of the
bifurcation point $(t^{*},p_i)$%
\begin{equation}
\label{107}v^A=f^A(v^1,t,\sigma ^2,\cdots ,\sigma ^k),\;\;\;\;\;A=2,3,...,n 
\end{equation}
with the partial derivatives%
$$
f_1^A=\frac{\partial v^A}{\partial v^1},\;\;\;f_t^A=\frac{\partial v^A}{%
\partial t},\;\;\;A=2,3,...,n. 
$$
Then, for $a=1,...,m-1$ we have%
$$
\phi ^a=\phi ^a(v^1,f^2(v^1,t,\vec \sigma ),...,f^m(v^1,t,\vec \sigma ),t,%
\vec \sigma )\equiv 0 
$$
which gives 
\begin{equation}
\label{108}\sum\limits_{A=2}^m\frac{\partial \phi ^a}{\partial v^A}f_1^A=-%
\frac{\partial \phi ^a}{\partial v^1},\;\;\;a=1,...,m-1 
\end{equation}
\begin{equation}
\label{109}\sum\limits_{A=2}^m\frac{\partial \phi ^a}{\partial v^A}f_t^A=-%
\frac{\partial \phi ^a}{\partial t},\;\;\;a=1,...,m-1 
\end{equation}
from which we can calculate the first order derivatives of $f^A$ : $f_1^A$
and $f_t^A$. Denoting the second order partial derivatives as%
$$
f_{11}^A=\frac{\partial ^2v^A}{(\partial v^1)^2},\;\;f_{1t}^A=\frac{\partial
^2v^A}{\partial v^1\partial t},\;\;\;f_{tt}^A=\frac{\partial ^2v^A}{\partial
t^2} 
$$
and differentiating (\ref{108}) with respect to $v^1$ and $t$ respectively,
we get 
\begin{equation}
\label{110}\sum\limits_{A=2}^m\phi _A^af_{11}^A=-\sum\limits_{A=2}^m[2\phi
_{A1}^af_1^A+\sum\limits_{B=2}^m(\phi _{AB}^af_1^B)f_1^A]-\phi
_{11}^a,\;\;\;a=1,2,...,m-1 
\end{equation}
\begin{equation}
\label{111}\sum\limits_{A=2}^m\phi _A^af_{1t}^A=-\sum\limits_{A=2}^m[\phi
_{At}^af_1^A+\phi _{A1}^af_t^A+\sum\limits_{B=2}^m(\phi
_{AB}^af_t^B)f_1^A]-\phi _{1t}^a,\;\;\;a=1,2,...,m-1. 
\end{equation}
And the differentiation of (\ref{109}) with respect to $t$ gives 
\begin{equation}
\label{112}\sum\limits_{A=2}^m\phi _A^af_{tt}^A=-\sum\limits_{A=2}^m[2\phi
_{At}^af_t^A+\sum\limits_{B=2}^m(\phi _{AB}^af_t^B)f_t^A]-\phi
_{tt}^a,\;\;\;a=1,2,...,m-1 
\end{equation}
where%
$$
\phi _{AB}^a=\frac{\partial ^2\phi ^a}{\partial v^A\partial v^B},\;\;\;\phi
_{At}^a=\frac{\partial ^2\phi ^a}{\partial v^A\partial t}. 
$$
The differentiation of (\ref{109}) with respect to $v^1$ gives the same
expression as (\ref{111}). If we use the Gaussian elimination method to the
three vectors at the right hands of the formulas (\ref{110}), (\ref{111})
and (\ref{112}), we can obtain the three partial derivatives $%
f_{11}^A,\;f_{1t}^A$ and $f_{tt}^A$. Notice that the three equations (\ref
{110}), (\ref{111}) and (\ref{112}) have the same coefficient matrix $J_1(%
\frac \phi v)$, which are assumed to be nonzero, and we should substitute
the values of the partial derivatives $f_1^A$ and $f_t^A$, which have been
calculated out in the former, into the right hands of the three equations.

The above discussions do not matter to the last component $\phi ^m(v^1,\cdot
\cdot \cdot ,v^m,t,\vec \sigma ).\,$ In order to find the different values
of $dv^1/dt$ at the bifurcation point, let us investigate the Taylor
expansion of $\phi ^m(v^1,\cdot \cdot \cdot ,v^m,t,\vec \sigma )$ in the
neighborhood of $(t^{*},p_i)$. Substituting the existing, but unknown,
dependency relationship (\ref{107}) into $\phi ^m(v^1,\cdot \cdot \cdot
,v^m,t,\vec \sigma )$, we get the function of two variables $v^1$ and $t$%
\begin{equation}
\label{113}F(t,v^1,\vec \sigma )=\phi ^m(v^1,f^2(v^1,t,\vec \sigma
),...,f^m(v^1,t,\vec \sigma ),t,\vec \sigma ) 
\end{equation}
which according to (\ref{phia}) must vanish at the bifurcation point 
\begin{equation}
\label{114}F(t^{*},p_i)=0. 
\end{equation}
From (\ref{113}), we can calculate the first order partial derivatives of $%
F(t,v^1,\vec \sigma )$ with respect to $v^1$ and $t$ respectively at the
bifurcation point $(t^{*},p_i)$%
\begin{equation}
\label{115}\frac{\partial F}{\partial v^1}=\phi _1^m+\sum\limits_{A=2}^m\phi
_A^mf_1^A,\;\;\;\frac{\partial F}{\partial t}=\phi
_t^m+\sum\limits_{A=2}^m\phi _A^mf_t^A. 
\end{equation}
Using (\ref{108}) and (\ref{109}), the first equation of (\ref{104}) is
expressed by%
$$
J(\frac \phi v)|_{(t^{*},p_i)}=\left| 
\begin{array}{cccc}
-\sum\limits_{A=2}^m\phi _A^1f_1^A & \phi _2^1 & \cdots & \phi _m^1 \\ 
-\sum\limits_{A=2}^m\phi _A^2f_1^A & \phi _2^2 & \cdots & \phi _m^2 \\ 
\vdots & \vdots & \cdot & \vdots \\ 
-\sum\limits_{A=2}^m\phi _A^{m-1}f_1^A & \phi _2^{m-1} & \cdots & \phi
_m^{m-1} \\ 
\phi _A^m & \phi _2^m & \cdots & \phi _m^m 
\end{array}
\right| _{(t^{*},p_i)}=0 
$$
which, by Cramer's rule, (\ref{106}) and (\ref{115}), can be rewritten as%
$$
J(\frac \phi v)|_{(t^{*},p_i)}=\left| 
\begin{array}{cccc}
0 & \phi _2^1 & \cdots & \phi _m^1 \\ 
0 & \phi _2^2 & \cdots & \phi _m^2 \\ 
\vdots & \vdots & \ddots & \vdots \\ 
0 & \phi _2^{m-1} & \cdots & \phi _m^{m-1} \\ 
\phi _1^m+\sum\limits_{A=2}^m\phi _A^mf_1^A & \phi _2^m & \cdots & \phi _m^m 
\end{array}
\right| _{(t^{*},p_i)} 
$$
$$
=\frac{\partial F}{\partial v^1}\det J_1(\frac \phi v)|_{(t^{*},p_i)}=0. 
$$
Since%
$$
\det J_1(\frac \phi v)|_{(t^{*},p_i)}\neq 0 
$$
which is our assumption, the above equation leads to 
\begin{equation}
\label{116}\frac{\partial F}{\partial v^1}|_{(t^{*},p_i)}=0. 
\end{equation}
With the same reasons, we can prove that 
\begin{equation}
\label{117}\frac{\partial F}{\partial t}|_{(t^{*},p_i)}=0. 
\end{equation}
The second order partial derivatives of the function $F(t,v^1,\vec \sigma )$
are easily to find out to be%
$$
\frac{\partial ^2F}{(\partial v^1)^2}=\phi _{11}^m+\sum\limits_{A=2}^m[2\phi
_{1A}^mf_1^A+\phi _A^mf_{11}^A+\sum\limits_{B=2}^m(\phi _{AB}^mf_1^B)f_1^A] 
$$
$$
\frac{\partial ^2F}{\partial v^1\partial t}=\phi
_{1t}^m+\sum\limits_{A=2}^m[\phi _{1A}^mf_t^A+\phi _{tA}^mf_1^A+\phi
_A^mf_{1t}^A+\sum\limits_{B=2}^m(\phi _{AB}^mf_t^B)f_1^A] 
$$
$$
\frac{\partial ^2F}{\partial t^2}=\phi _{tt}^m+\sum\limits_{A=2}^m[2\phi
_{At}^mf_t^A+\phi _A^mf_{tt}^A+\sum\limits_{B=2}^m(\phi _{AB}^mf_t^B)f_t^A] 
$$
which at $(t^{*},p_i)$ are denoted by 
\begin{equation}
\label{118}A=\frac{\partial ^2F}{(\partial v^1)^2}\mid _{(t^{*},p_i)},\quad
\quad B=\frac{\partial ^2F}{\partial v^1\partial t}\mid _{(t^{*},p_i)},\
\quad \quad C=\frac{\partial ^2F}{\partial t^2}\mid _{(t^{*},p_i)}. 
\end{equation}
Then, by virtue of (\ref{114}), (\ref{116}), (\ref{117}) and (\ref{118}),
the Taylor expansion of $F(t,v^1,\vec \sigma )$ in the neighborhood of the
bifurcation point $(t^{*},p_i)$ can be expressed as 
\begin{equation}
\label{119}F(t,v^1,\vec \sigma )=\frac 12%
A(v^1-p_i^1)^2+B(v^1-p_i^1)(t-t^{*})+\frac 12C(t-t^{*})^2 
\end{equation}
which is the expression of $\phi ^m(v^1,\cdots ,v^m,t,\vec \sigma )$ in the
neighborhood of $(t^{*},p_i)$. The expression (\ref{119}) shows that at the
bifurcation point $(t^{*},p_i)$ 
\begin{equation}
\label{120}A(v^1-p_i^1)^2+2B(v^1-p_i^1)(t-t^{*})+C(t-t^{*})^2=0. 
\end{equation}
Dividing (\ref{120}) by $(v^1-p_i^1)^2$ or $(t-t^{*})^2$, and taking the
limit $t\rightarrow t^{*}$ as well as $v^1\rightarrow p_i^1$ respectively,
we get two equations 
\begin{equation}
\label{121}C(\frac{dt}{dv^1})^2+2B\frac{dt}{dv^1}+A=0. 
\end{equation}
and 
\begin{equation}
\label{122}A(\frac{dv^1}{dt})^2+2B\frac{dv^1}{dt}+C=0. 
\end{equation}
So we get the different directions of the branch curves at the bifurcation
point from the solutions of (\ref{121}) or (\ref{122}).

In order to determine the branches directions of the remainder variables, we
will use the relations simply%
$$
dv^A=f_1^Adv^1+f_t^Adt,\;\;\;\;\;A=2,3,...,n 
$$
where the partial derivative coefficients $f_1^A$ and $f_t^A$ have given in (%
\ref{108}) and (\ref{109}). Then, respectively%
$$
\frac{dv^A}{dv^1}=f_1^A+f_t^A\frac{dt}{dv^1} 
$$
or 
\begin{equation}
\label{126}\frac{dv^A}{dt}=f_1^A\frac{dv^1}{dt}+f_t^A. 
\end{equation}
where partial derivative coefficients $f_1^A$ and $f_t^A$ are given by (\ref
{108}) and (\ref{109}). From this relations we find that the values of $%
dv^A/dt$ at the bifurcation point $(t^{*},z_i)$ are also possibly different
because (\ref{122}) may give different values of $dv^1/dt$. The above solutions
reveal the evolution of the topological defects. Besides the encountering of
the defects, i.e. two defects encounter and then depart at the bifurcation point
along different branch curves, it also includes splitting and merging of defects.
When a multicharged defects moves through the bifurcation point, it may split into
several defects along these two different branch curves. On the contrary, several defects
can merge into one defect at the bifurcation point. The identical conservation of
the topological charge shows the sum of the topological charge of final defects
must be equal to that of the initial defects at the bifurcation point,
i.e.,
\begin{equation}
\label{chargeII}\sum_f\beta _{j_f}\eta _{j_f}=\sum_i\beta _{j_i}\eta _{j_i} 
\end{equation}
for fixed $j$. Furthermore, from above studies, we see that the generation,
annihilation and bifurcation of defects are not gradual changes, but start
at a critical value of arguments, i.e. a sudden change.

\vskip0.5cm
{\bf C. Branch process at a higher degenerated point}

\vskip0.5cm
In the following, let us discuss the branch process at a higher degenerated
point. In the above subsection, we have analysised the case that the rank of
the Jacobian matrix $[\partial \phi /\partial v]$ of the equation (\ref{zero}%
) is $m-1$. In this section, we consider the case that the rank of the
Jacobian matrix is $m-2$ (for the case that the rank of the matrix $%
[\partial \phi /\partial v]$ is lower than $m-2$, the discussion is in the
same way). Let the $(m-2)\times (m-2)$ submatrix $J_2(\frac \phi v)$ of the
Jacobian matrix $[\partial \phi /\partial v]$ be%
$$
J_2(\frac \phi v)=\left( 
\begin{array}{cccc}
\phi _3^1 & \phi _4^1 & \cdots & \phi _m^1 \\ 
\phi _3^2 & \phi _4^2 & \cdots & \phi _m^2 \\ 
\vdots & \vdots & \ddots & \vdots \\ 
\phi _3^{m-2} & \phi _4^{m-2} & \cdots & \phi _m^{m-2} 
\end{array}
\right) 
$$
and suppose that $\det J_2(\frac \phi v)|_{(t^{*},p_i)}\neq 0.$ With the
same reasons of obtaining (\ref{107}), we can have the function relations 
\begin{equation}
\label{127}v^A=f^A(v^1,v^2,t,\vec \sigma ),\;\;\;\;\;A=3,4,...,m. 
\end{equation}
For the partial derivatives $f_1^A$, $f_2^A$ and $f_t^A$, we can easily
derive the system similar to the equations (\ref{108}) and (\ref{109}), in
which the three terms at the right hand of can be figured out at the same
time. In order to determine the 2--order partial derivatives $f_{11}^A$, $%
f_{12}^A$, $f_{1t}^A$, $f_{22}^A$, $f_{2t}^A$ and $f_{tt}^A$, we can use the
equations similar to (\ref{110}), (\ref{111}) and (\ref{112}). Substituting
the relations (\ref{127}) into the last two equations of the system (\ref
{phia}), we have the following two equations with respect to the arguments $%
v^1,\,\,v^2,\,\,t,\vec \sigma $%
\begin{equation}
\label{bifa46}\left\{ 
\begin{array}{l}
F_1(v^1,v^2,t, 
\vec \sigma )=\phi ^{m-1}(v^1,v^2,f^3(v^1,v^2,t,\vec \sigma ),\cdots
,f^m(v^1,v^2,t,\vec \sigma ),t,\vec \sigma )=0 \\ F_2(v^1,v^2,t,\vec \sigma
)=\phi ^m(v^1,v^2,f^3(v^1,v^2,t,\vec \sigma ),\cdots ,f^m(v^1,v^2,t,\vec 
\sigma ),t,\vec \sigma )=0. 
\end{array}
\right. 
\end{equation}
Calculating the partial derivatives of the function $F_1$ and $F_2$ with
respect to $v^1$, $v^2$ and $t$, taking notice of (\ref{127}) and using six
similar expressions to (\ref{116}) and (\ref{117}), i.e. 
\begin{equation}
\label{bifa48}\frac{\partial F_j}{\partial v^1}\mid _{(t^{*},p_i)}=0,\ \quad
\quad \frac{\partial F_j}{\partial v^2}\mid _{(t^{*},p_i)}=0,\ \quad \quad 
\frac{\partial F_j}{\partial t}\mid _{(t^{*},p_i)}=0,\ \quad \quad j=1,2, 
\end{equation}
we have the following forms of Taylor expressions of $F_1$ and $F_2$ in the
neighborhood of $(t^{*},p_i)$ 
$$
F_j(v^1,v^2,t,\vec \sigma )\approx
A_{j1}(v^1-p_i^1)^2+A_{j2}(v^1-p_i^1)(v^2-p_i^2)+A_{j3}(v^1-p_i^1) 
$$
$$
(t-t^{*})+A_{j4}(v^2-p_i^2)^2+A_{j5}(v^2-p_i^2)(t-t^{*})+A_{j6}(t-t^{*})^2=0 
$$
\begin{equation}
\label{bifa49}j=1,2. 
\end{equation}
In the case of $A_{j1}\neq 0,A_{j4}\neq 0$, by dividing (\ref{bifa49}) by $%
(t-t^{*})^2$ and taking the limit $t\rightarrow t^{*}$, we obtain two
quadratic equations of $\frac{dv^1}{dt}$ and $\frac{dv^2}{dt}$ 
\begin{equation}
\label{bifa50}A_{j1}(\frac{dv^1}{dt})^2+A_{j2}\frac{dv^1}{dt}\frac{dv^2}{dt}%
+A_{j3}\frac{dv^1}{dt}+A_{j4}(\frac{dv^2}{dt})^2+A_{j5}\frac{dv^2}{dt}%
+A_{j6}=0 
\end{equation}
$$
j=1,2. 
$$
Eliminating the variable $dv^1/dt$, we obtain a equation of $dv^2/dt$ in the
form of a determinant 
\begin{equation}
\label{bifa51}\left| 
\begin{array}{cccc}
A_{11} & A_{12}Q+A_{23} & A_{14}Q^2+A_{15}Q+A_{16} & 0 \\ 
0 & A_{11} & A_{12}Q+A_{13} & A_{14}Q^2+A_{15}Q+A_{16} \\ 
A_{21} & A_{22}Q+A_{23} & A_{24}Q^2+A_{25}Q+A_{26} & 0 \\ 
0 & A_{21} & A_{22}Q+A_{23} & A_{24}Q^2+A_{25}Q+A_{26} 
\end{array}
\right| =0 
\end{equation}
where $Q=dv^2/dt$, which is a $4th$ order equation of $dv^2/dt$ 
\begin{equation}
\label{bifa52}a_0(\frac{dv^2}{dt})^4+a_1(\frac{dv^2}{dt})^3+a_2(\frac{dv^2}{%
dt})^2+a_3(\frac{dv^2}{dt})+a_4=0. 
\end{equation}
Therefore we get different directions at the bifurcation point corresponding
to different branch curves. The number of different branch curves is four at
most. If the degree of degeneracy of the matrix $[\frac{\partial \phi }{%
\partial v}]$ is more higher, i.e. the rank of the matrix $[\frac{\partial
\phi }{\partial v}]$ is more lower than the present $(m-2)$ case, the
procedure of deduction will be more complicate. In general supposing the
rank of the matrix $[\frac{\partial \phi }{\partial x}]$ be $(m-s)$, the
number of the possible different directions of the branch curves is $2^s$ at
most. Comparing with the above subsection, the solutions in the subsection also
reveal encountering, splitting and merging of the defects along more directions.

At the end of this section, we conclude that there exist crucial cases of
branch processes in our topological defect theory. This means that a
topological defect, at the bifurcation point, may split into several
topological defects along $2^s$ different branch curves with
different charges. Since the topological current is a conserved current, the
total quantum number of the spliting topological defects must precisely
equal to the topological charge of the original defect i.e. 
$$
\sum\limits_{j=1}^{2^s}\beta _{i_j}\eta _{i_j}=\beta _i\eta _i
$$
for fixed $i$, $\beta_{i_j}\eta_{i_j}$ stands for the total topological
charge of defects along the $j$-th branch curve at the bifurcation point
$p_i$. This can be looked upon as the topological reason of the
defect splitting. Here we should point out that such splitting is a
stochastic process, the sole restriction of this process is just the
conservation of the topological charge of the topological defects during
this splitting process.

\end{document}